\documentclass[journal]{IEEEtranTIE}
\usepackage{graphicx}
\usepackage{cite}
\usepackage{picinpar}
\usepackage{amsmath}
\usepackage{amssymb}
\usepackage{url}
\usepackage{flushend}
\usepackage[latin1]{inputenc}
\usepackage{colortbl}
\usepackage{soul}
\usepackage{multirow}
\usepackage{pifont}
\usepackage{color}
\usepackage{alltt}
\usepackage[hidelinks]{hyperref}
\usepackage{enumerate}
\usepackage{siunitx}
\usepackage{breakurl}
\usepackage{epstopdf}
\usepackage{pbox}

\hypersetup{draft}

\begin{document}
\title{Automatic LiDAR Extrinsic Calibration System using Photodetector and Planar Board for Large-scale Applications}

\author{
	\vskip 1em
	
	Ji-Hwan You, Seon Taek Oh, Jae-Eun Park,
	Azim Eskandarian, \emph{IEEE Senior Member},
	\\ and Young-Keun Kim

	\thanks{
	

		Seontake Oh, Ji-Hwan You, Jae-Eun Park are with the School of Mechanical and Control Engineering, Handong Global University, Pohang, Rep. Korea 
		
		Azim Eskandarian is with the department  of Mechanical Engineering, Blacksburg, Virginia Tech, VA, USA 
		
		Young-Keun Kim is with the School of Mechanical and Control Engineering, Handong Global University, Pohang, Rep. Korea (corresponding author e-mail:ykkim@handong.edu). 

	}
}

\maketitle
	
\begin{abstract}
This paper presents a novel automatic calibration system to estimate the extrinsic parameters of LiDAR mounted on a mobile platform for sensor misalignment inspection in the large-scale production of highly automated vehicles.
To obtain subdegree and subcentimeter accuracy levels of extrinsic calibration, this study proposed a new concept of a target board with embedded photodetector arrays, named the PD-target system, to find the precise position of the correspondence laser beams on the target surface. 
Furthermore, the proposed system requires only the simple design of the target board at the fixed pose in a close range to be readily applicable in the automobile manufacturing environment. 
The experimental evaluation of the proposed system on low-resolution LiDAR showed that the LiDAR offset pose can be estimated within 0.1 degree and 3 mm levels of precision.
The high accuracy and simplicity of the proposed calibration system make it practical for large-scale applications for the reliability and safety of autonomous systems.
\end{abstract}


{}
\definecolor{limegreen}{rgb}{0.2, 0.8, 0.2}
\definecolor{forestgreen}{rgb}{0.13, 0.55, 0.13}
\definecolor{greenhtml}{rgb}{0.0, 0.5, 0.0}

\section{Introduction}
\IEEEPARstart{T}{he} LiDAR (light detection and ranging) is mounted on mobile platforms, for locating multiple targets and obstacles \cite{ Song2016, Kang2012, Ramasamy2016} for tasks such as collision avoidance and path planning \cite{Zeng2018}, and generating 3D maps of the surroundings for localization\cite{Zhang2020,Wang2017}.
To generate accurate reconstruction and interpretation of the surroundings, precise assembly of LiDAR on mobile platforms is critical. 
An alignment offset as low as a few degrees could cause a significant error in predicting the position of obstacles relative to the ego-vehicle and can result in dangerous situations. 

Thus, for large-scale production of vehicles with attached LiDAR for a high level of ADAS, the sensor misalignment inspection is necessary because an unexpected offset in the sensor assembly reduces the reliability and safety of the autonomous system.
The sensor misalignment can be inspected by the extrinsic parameter calibration that estimates the pose of the sensor frame with respect to the reference frame of the mobile platform.

Therefore, this paper aims to design an automatic and accurate LiDAR calibration system for practical application in the vehicle assembly process in the automobile industry.


Most studies on LiDAR extrinsic calibration are focused on solving for the geometric relationship with respect to a camera. In contrast, there are only a few studies on estimating the LiDAR misalignment with respect to the vehicle body frame, as explained in Section II.
These studies of LiDAR extrinsic calibration use the flat ground plane and a long pole from a series of poses\cite{Zhu2013}, the ground as the target object at various poses\cite{Zaiter2019} or multiple planar objects(walls and floor)\cite{Atanacio-Jimenez2011} to estimate the rigid transformation between the LiDAR and the target objects.


However, considering the actual environment of the automobile manufacturing process, the inspection area is confined to a limited space of a few meters that includes other machinery for various assembly tasks, such as rails on the ground. 

Therefore, to overcome these limitations, this paper proposes a new extrinsic calibration method that uses a single calibrating target, at a close range and requires only one target pose. Most importantly, the misalignment pose is estimated with an accuracy within a few millimeters and at the subdegree level even for a low-resolution mobile LiDAR.

This paper designed a novel automatic calibration system composed of a simple planar target with embedded photodetector arrays, named as the PD-target system.
Modules of NIR photodetector 1D arrays, are arranged near the corners of a planar mid-sized target, to detect the the position of the beam spot on the target surface. 
The positions of the corresponding beam points measured from the LiDAR point cloud data and the target photodetectors are used for the initial pose estimation. Then, an iterative nonlinear optimization method is applied to obtain the final pose of the LiDAR with respect to the target body frame, which, without loss of generality, is assumed to be the vehicle body frame. 

The contributions of this paper can be summarized as follows. 
First, the proposed system is a novel calibration method that combines a target board with photodetector arrays (PD-target) that has not been proposed in other literature. 
Second, the system can estimate the sensor alignment within a high level of subdegree and subcentimeter accuracy with the help of NIR photodetector sensors. 
Last, this system requires only one pose of a simple target board in a close range of few meters that can be applicable in a large-scale automobile process.



\section{Related Work}
LiDAR calibration can be classified as intrinsic parameter calibration and extrinsic parameter calibration. 
The intrinsic calibration refers to optimizing the internal parameters to minimize the systematic error of the LiDAR measurements due to the optical parts and conversion from raw data to 3D points\cite{Atanacio-Jimenez2011}, \cite{Muhammad2010} \cite{Glennie2010}. However, this paper focused on the extrinsic calibration of LiDAR.


Most research on LiDAR extrinsic calibration are focused on estimating the geometric relationship of the LiDAR with respect to the camera.
The extrinsic parameters are estimated by extracting the 3D-2D correspondences from the point cloud of 3D data and 2D image pixels using one or multiple planar targets\cite{geiger2012automatic, kim2019extrinsic}, of various such as circular \cite{fremont2008extrinsic, alismail2012automatic}, trihedral \cite{gong20133d}  and 3D cubic object\cite{pusztai2017accurate}.
A common method is using a chessboard at multiple poses to determine the correspondence of the planes and lines and applying an optimization method \cite{zhang2004extrinsic, verma2019automatic}, \cite{Mirzaei2012}.

In contrast to the research on the LiDAR extrinsic calibration with a camera, the LiDAR misalignment with respect to the vehicle body frame has not been investigated to that extent.

Several studies proposed measuring systems that estimate the relative 6-DOF pose between a planar target and the a sensor system composed of 1D or 2D laser scanners
\cite{kim2012developing, kim2014design, kim2015portable, kim2014structural, kim2014note}.
These systems detect the feature beam points by using a camera or using the edges of the planar targets.
A method of calibrating multiple 2D laser scanners with respect to the vehicle body was proposed in \cite{Underwood2010} that used a vertical pole on flat ground.

An extrinsic calibration for a 3D LiDAR was proposed by \cite{Zhu2013} that  first computes the pitch and roll angles from the flat ground plane and then computes the navigation angle by using a long pole from a series of poses with on-vehicle IMUs and GPS. 
A similar method was introduced to estimate the extrinsic LiDAR parameters with respect to the reference ground in \cite{Zaiter2019}. Here, it used the best-fit plane model of the ground and least squares conic algorithm to estimate for the Euler angles and sensor elevation above the ground.
Other extrinsic parameter methods of a 3D LiDAR used a calibration object consisting of four walls and the floor\cite{Atanacio-Jimenez2011}. After estimating the sensor intrinsic parameters, the extrinsic rigid transformation between the LiDAR and the target objects was estimated with an optimization method. 



Since these methods have limitations to satisfy the aforementioned large-scale application, this paper proposes a new extrinsic calibration method using the PD-target system, which is composed of a planar target with photodetector arrays.

\section{Analysis of LiDAR Measurement Error}

\subsection{Resolution of 3D measurement}  
The LiDAR model used in this paper is Velodyne VLP-16, which has 16 vertical channels of LEDs emitting 903 nm near-infrared (NIR) wavelength.
The firing of beam pulses is a series of a short bursts with a period of 2.3 $\mu sec$. 
Therefore, the minimum speed of ADC for capturing a LiDAR pulse should be at least 1 MHz.

\begin{figure}
    \centering
    \includegraphics[width=0.9\columnwidth]{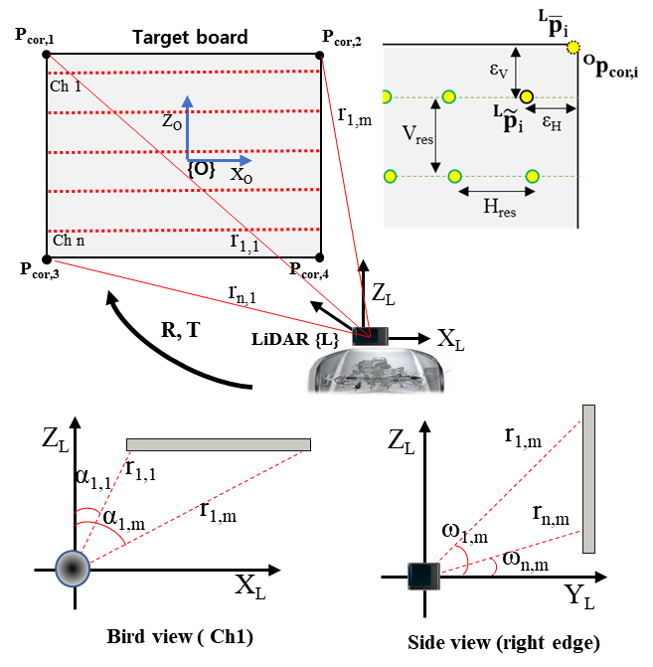}
    \caption{\label{Fig_TargetboardROI} The geometric relationship of 3D rotation (R) and translation (T) between the target body frame and the LiDAR frame are shown. The number of scan channels of LiDAR is n, and each beam with index (i,j) is measured in terms of the range (r), azimuth($\alpha$) and vertical angle ($\omega$).
    }
    \end{figure}

From the raw LiDAR data, the beams on the target board can be defined to be bounded by the vertical angle ranging from $\omega_1$ to $\omega_n$ and the azimuth angle from $\alpha_1$  to $\alpha_m$, where the values of $n$ and $m$ are the maximum numbers of scan channels and the number of beams per channel on the given target board dimension, respectively, as described in Fig. \ref{Fig_TargetboardROI}.
VLP-16 has an azimuth and vertical angular resolution of 0.2 and 2 degrees, respectively, and has the $H_{res}$ of 9 mm and $V_{res}$ of 87 mm, on the target board located at 2.5 m of distance.

\subsection{Uncertainty of 3D measurement} 
The principle in estimating the relative pose between the LiDAR and the target is to find the correspondence of feature points, ${}^OP$ and ${}^LP$, which are measured from the two different coordinate frames of the LiDAR body, $\{L\}$, and the target system, $\{O\}$, respectively, as shown in Fig. \ref{Fig_TargetboardROI}. The relationship between the two groups of correspondence feature points, ${}^OP$ and ${}^LP$, describes the orientation matrix $\bf{R}$ and the translation vector $\bf{T}$ of the LiDAR with respect to the reference target board.


    \begin{equation}
    {}^{\bf{O}}{\bf{P}} = {}_{\bf{L}}^{\bf{O}}{\bf{M}}{}^{\bf{L}}{\bf{P}} = [{}_{\bf{L}}^{\bf{O}}{\bf{R}}|{}_{\bf{L}}^{\bf{O}}{\bf{T}}]{}^{\bf{L}}{\bf{P}}
    \label{eq:pose}
    \end{equation}

Typically, corners and edges are used as the features of a target for LiDAR calibration. Let the corners of the target board, $oP_{cor,i}$, be the features that need to be detected by LiDAR, as indicated in Fig. \ref{Fig_TargetboardROI}. For the ideal measurement of that feature point, $ {}^{\bf{L}}{\overline {\bf{p}} _{i,k}}$, the LiDAR should project a beam spot directly on that corner. 

    \begin{equation}
    {}^{\bf{L}}{\overline {\bf{p}} _{i,k}} = ({r_{i,k}})\left[ {\begin{array}{*{20}{c}}
    {\cos ({\omega _{i,k}})\sin ({\alpha _{i,k}})}\\
    {\cos ({\omega _{i,k}})\cos ({\alpha _{i,k}})}\\
    {\sin ({\alpha _{i,k}})}
    \end{array}} \right]
    \end{equation}

However, the actual position of the beam closest to the corner, ${}^{\bf{L}}{\widetilde {\bf{p}}}$, is restricted by the sensor resolution 

    \begin{equation}
    {}^{\bf{L}}{\widetilde {\bf{p}}_{i,k}} = ({\overline r _{i,k}} + {\varepsilon _k})\left[ {\begin{array}{*{20}{c}}
    {\cos ({{\overline \omega  }_{i,k}} + \delta {\omega _k})\sin ({{\overline \alpha  }_{i,k}} + \delta {\alpha _k})}\\
    {\cos ({{\overline \omega  }_{i,k}} + \delta {\omega _k})\cos ({{\overline \alpha  }_{i,k}} + \delta {\alpha _k})}\\
    {\sin ({{\overline \alpha  }_{i,k}} + \delta {\alpha _k})}
    \end{array}} \right]
    \end{equation}where $\epsilon _k$ is the depth measurement noise, and $\delta \omega$ and $\delta \alpha$ are the uncertainty of the azimuth and the vertical angle resolution, respectively. The subscript $i$ is the point closest to the $i^{th}$ corner, and $k$ is the scan number. 

The deviation of the beam point from the corner of the target surface is bounded by the azimuth and vertical angle resolution as

    \begin{equation}
        {}^{\bf{L}}{\overline {\bf{p}} _{i,k}} - {}^{\bf{L}}{\widetilde {\bf{p}}_{i,k}} \le {{\bf{\varepsilon }}_{i,k}} = \left[ {\begin{array}{*{20}{c}}
        {{r_{i,k}}\tan \delta {\alpha _k}}\\
        0\\
        {{r_{i,k}}\tan \delta {\omega _k}}
        \end{array}} \right]
    \end{equation}


This correspondence uncertainty needs to be minimized to obtain higher accuracy in the relative pose estimation. Since the resolution of the given LiDAR cannot be modified unless a higher performance LiDAR replaces it, the feature on the target board needs to be shifted from the corners (${}^OP_{cor}$) to the position where beams are actually projected, (${}^OP_{i}$). 

However, this introduces a technical problem of measuring the position of ${}^OP_{i}$ by the target board. 
Thus, this paper proposes a method to measure the position ${}^OP_{i}$ with high precision by  using photodetectors attached on the target surface.



\section{Design of PD-Target System}

\subsection{Photodetector array}
\begin{figure}[!t]
    \centering
    \includegraphics[width=0.9\columnwidth]{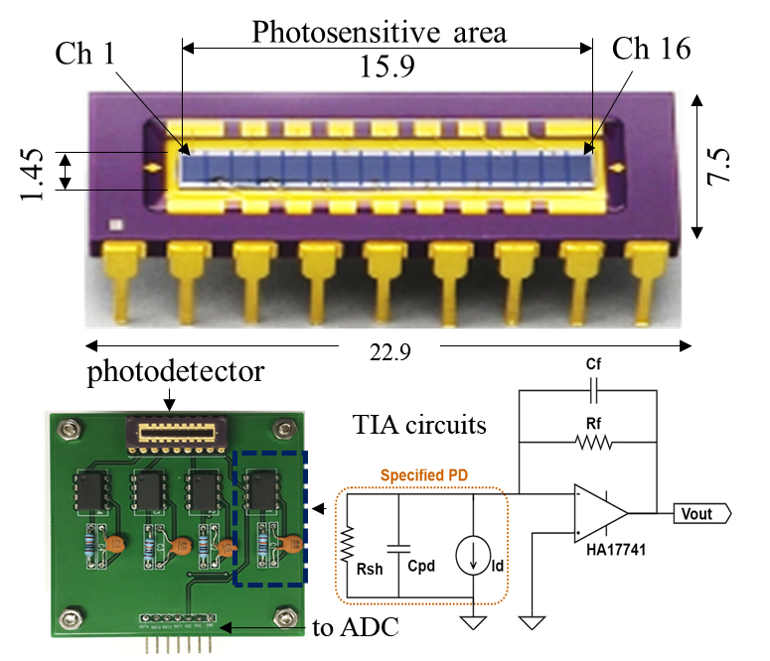}
    \caption{Photodetector 1-D array of 16 diode elements (Hamamatsu S4111) and the PCB of the signal processing circuit designed in this study
    }\label{Fig_PD}
    \end{figure}

Based on the analysis of the laser beams in Section III., the appropriate photodetector should be sensitive to the NIR spectrum and have a bandwidth higher than 1 MHz. In addition, the active sensor area of the photodetector should be large enough to capture the beam spot diameter.
This paper selected a photodetector 1-D array of 16 diode elements (Hamamatsu S4111), which is sensitive from the UV to the NIR spectrum. 
The active sensor area is 16 mm by 1.45 mm and the overall dimension, is shown in Fig. \ref{Fig_PD}.


\subsection{PD-Target board configuration}
The photodetector(PD) 1-D arrays are positioned on the target board to capture the LiDAR beam points, which are near the target edges and on the first and the last vertical laser scans.
Since each photodiode is a 1D array, it can detect the center position of a single beam spot in its own coordinate frame, $\{D\}$. The beams  photodiodes are referred to as the key beam points throughout the paper. 

    
Then, the position of the beam in the target coordinate frame $\{O\}$ can be easily be converted from $\{D\}$ with the matrix $P_{offset}=[X_{pd,offset},Z_{pd,offset}]$ that describes the location of each photodetector attachment from the center of the board. 
    \begin{equation}
    {}^{\bf{O}}{\bf{P}} = {}^D{\bf{P}} - {{\bf{P}}_{offset}}
    \end{equation}

The 1D PD could be attached to the target surface at any angle. However, we considered the horizontal and vertical orientations and compared the performance of these two configurations.

\subsection{Signal Processing Circuit}
The photodetector diode generates an electric current when the LiDAR beam is projected on it. 
The output of the PD should be pre-amplified voltage to be used as the source for the ADC that is connected to the main processor. Thus, the trans-impedance amplifier (TIA) circuit is designed to convert the current to voltage and amplify the output to the usable level. The designed TIA circuit consists of an OP-amplifier, a feedback resistor, and a capacitor, as shown in Fig. \ref{Fig_PD}.

  \begin{table}[t]
    \caption{OP-AMP and photodetector parameters}
    \label{tab:opamp}
    \center
      \begin{tabular*}{\linewidth}{c|cc}
            \hline
            Symbol& Description& Value\\
            \hline
            GBWP&	Op-amp gain bandwidth product& 	1[MHz]\\
            $A_{OL}$&	Op-amp DC open-loop gain&	106[dB]\\
            $C_{i,amp}$&	Op-amp input capacitance&	1.4[pF]\\
            $C_{pd}$&	Photodiode junction capacitance&	200[pF]\\
            $R_{sh}$&	Photodiode shunt resistance&	250[G$\Omega$]\\
            \hline
        \end{tabular*}
    \end{table}


The value of the feedback resistor, $R_f$, is determined from the ratio of the maximum photodiode current and the preferred range of the output voltage. The electric current of the PD generated by the laser photons from 2.5 m is expected to be a maximum of 100 $\mu A$. 

Targeting the output voltage to be within the range of 0 V to 10 V, the Rf value was designed to be 100 k$\Omega$.

The voltage out of the TIA circuit can be expressed as a first-order system as

    \begin{equation}
    V_{out} = \frac{{{I_d}{R_f}}}{{({R_f}\;{C_f}\;s + 1)\;}}\
    \end{equation}

However, to avoid any unwanted oscillation in the signal, the stability of the circuit needs to be analyzed with the noise gain, which is the transfer function of the 1st order system of one zero and pole, composed of all resistors and capacitors in the amplifier and the photodetector.

    \begin{equation}
    {G_{noise}}\;(s) = \frac{{({R_f} + {R_{sh}}\;)}}{{{R_{sh}}}}\frac{{\left( {\frac{{{R_f}{R_{sh}}}}{{{R_f} + {R_{sh}}\;}}} \right)({C_f} + {C_{in}})s + 1}}{{{R_f}\;{C_f}\;s + 1}}
    \end{equation} 
    where $C_{in}$ is the combination of capacitors at the input of the amplifier. All the parameters except the variable $R_f$ and $C_f$ are predetermined by the parts specification.

The oscillation or the stability of the signal can be predicted with the circuit quality factor (Q), which can be calculated from the TIA circuit phase margin\cite{Fuada2017}.

    \begin{equation}
    Q = {\left( {{{\left( {\frac{1}{{{{(\tan \,\varphi )}^2}}}\; + 0.5} \right)}^2} - 0.25} \right)^{\frac{1}{4}}}
    \end{equation}



A Q-factor value below 0.5 causes the system to be overdamped to avoid any oscillation in the output voltage. In this application,  the feedback  capacitance of 68 pF was selected and resulted in a Q-value of 0.47 that yielded an overdamped system. 

Circuit simulation with the designed TIA circuit is conducted by applying 100 ms (10 Hz) of laser pulses with different power intensities to generate the current input ranging from 100 to 400 $\mu A$ to the circuit. 

\begin{figure}[!t]
    \centering
    \includegraphics[width=1\columnwidth]{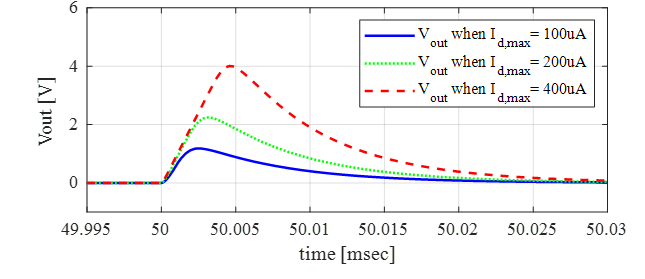}
    \caption{ Simulation of the designed TIA circuit with an input signal of 10-Hz laser pulses with different power intensities that generate current of 100 to 400 $\mu A$ from the photodetector 
    }\label{Fig_CircuitSim}
    \end{figure}

 The results in Fig. \ref{Fig_CircuitSim}showed that the output voltage can be generated as high as 4V when  the current of 400 $\mu A$ is applied by laser beam.Moreover, there is no oscillation in the output signals, as required. The circuitry of TIA was designed on a PCB that can amplify up to 4 diode channels of a PD array, as shown in Fig. \ref{Fig_PD}. 




The sampling rate for each ADC channel should be higher than 2 MHz to capture laser pulses of the 2.2-$\mu sec$ period. In addition, all the ADC channels must be concurrent to capture all the output signals of the PD simultaneously. 
 Limited by the availability of ADC channels and DAQ boards, only two PDs and four out of sixteen diodes from each PD were sampled simultaneously.

\section{Algorithm of LiDAR Extrinsic Calibration}

\subsection{Preprocessing}
From the whole point cloud of one LiDAR scan, the target board is segmented as the region of interest (ROI).
A Euclidean clustering algorithm is applied to segment the planar target automatically from the background.
With the point cloud data of the ROI, the best-fit plane model of the target is derived using a least squares method. 
Then, the ROI beam points are projected onto the fit plane surface to minimize the sensor measurement error.

    

\subsection{Relative pose of LiDAR}

The relative 6-DOF pose of the LiDAR from the reference target frame can be estimated by solving for the transformation matrix ${}_L^OM(\phi, \theta, \psi, \Delta x,\Delta y,\Delta z)$, as mentioned in Eq. \ref{eq:pose}. 
To solve for the transformation matrix ${}_{\bf{L}}^{\bf{O}}{\bf{M}}$, the 3D positions of feature beam points measured by the target body frame (${}^{\bf{O}}{\bf{p_i}}$) and measured in the LiDAR frame (${}^{\bf{L}}{\bf{p_i}}$) are necessary. With the proposed PD-target system, Eq. \ref{eq:pose} is modified to include the photodetector coordinate frame, ${D}$.

    \begin{equation}
    {}^D{\bf{P}} - {{\bf{P}}_{offset}}
    ={}_{\bf{L}}^{\bf{O}}{\bf{M}}{}^{\bf{L}}{\bf{P}} = [{}_{\bf{L}}^{\bf{O}}{\bf{R}}|{}_{\bf{L}}^{\bf{O}}{\bf{T}}]{}^{\bf{L}}{\bf{P}}
    \end{equation}
    
    where
    \begin{equation*}
    {}_{\bf{L}}^{\bf{O}}{\bf{R}} = \bf{R_z}(\phi)\bf{R_y}(\theta)\bf{R_x}(\psi),~~    {}_{\bf{L}}^{\bf{O}}{\bf{T}} = [ \Delta X, \Delta Y, \Delta Z]^T
    \end{equation*}
    
    

The process of obtaining the position of the correspondence feature points,${}^D{\bf{P}}$and ${}^L{\bf{P}}$ are described in the following section. Note that since the signal processing for each PD alignment are similar, only the horizontal aligned photodetectors are explained in the section.

\subsection{Estimating the position of the beam center}
\begin{figure}[!t]
    \centering
    \includegraphics[width=1\columnwidth]{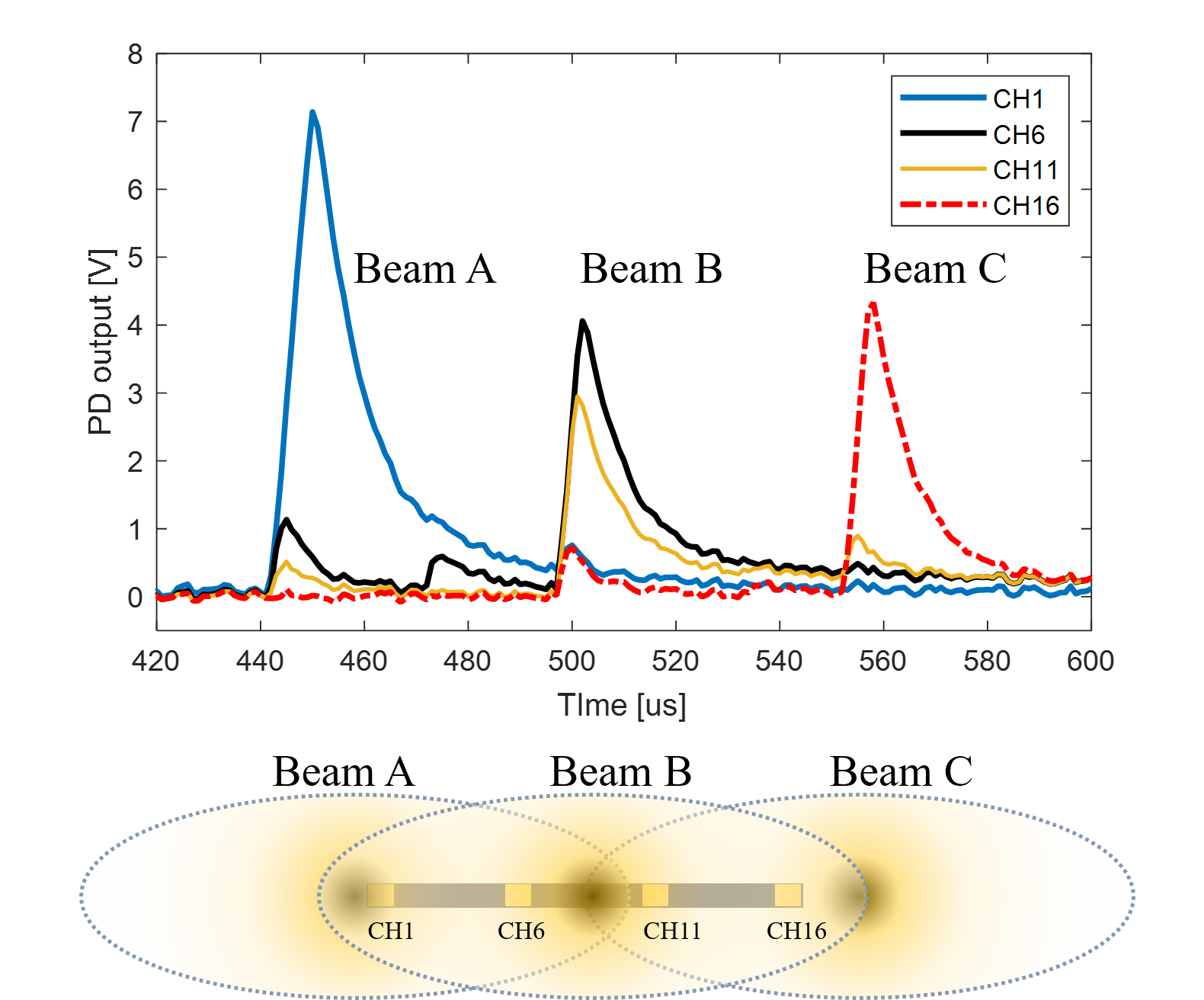}
    \caption{Three beams (A to C)  projected on the horizontally aligned photodetector in consequent time that generate signals from each diode 
    }\label{Fig_PD_H_response}
    \end{figure}

 For each LiDAR scan, there would be at most three beams projected on the same photodetector in the horizontal alignment, as shown in Fig. \ref{Fig_PD_H_response}. With the given configuration of the target board at a distance of 2.5 meters, the diameter of a beam spot would be 19.6 mm. The beams on the photosensitive area overlap with an interval of 9.7 mm but at different times with 55-$\mu sec$ period. 
    
The diode channel closest to the beam center outputs the highest peak level because the beam power intensity is approximately distributed as a bell-shaped. The first set of peaks is generated by $beam A$, and $CH1$ showed a higher level than the others because it is the closest to the beam center. After 55 $\mu s$ from the first peaks, $beam B$ is projected on the PD closest to $CH11$, which could be inferred by comparing the peak level among other channel outputs.  Thus, by analyzing the distribution of the peak values for multiple beam projections, the position and the time of the beam scanning over the PD can be derived.

\begin{figure}[]
    \centering
    \includegraphics[width=1\columnwidth]{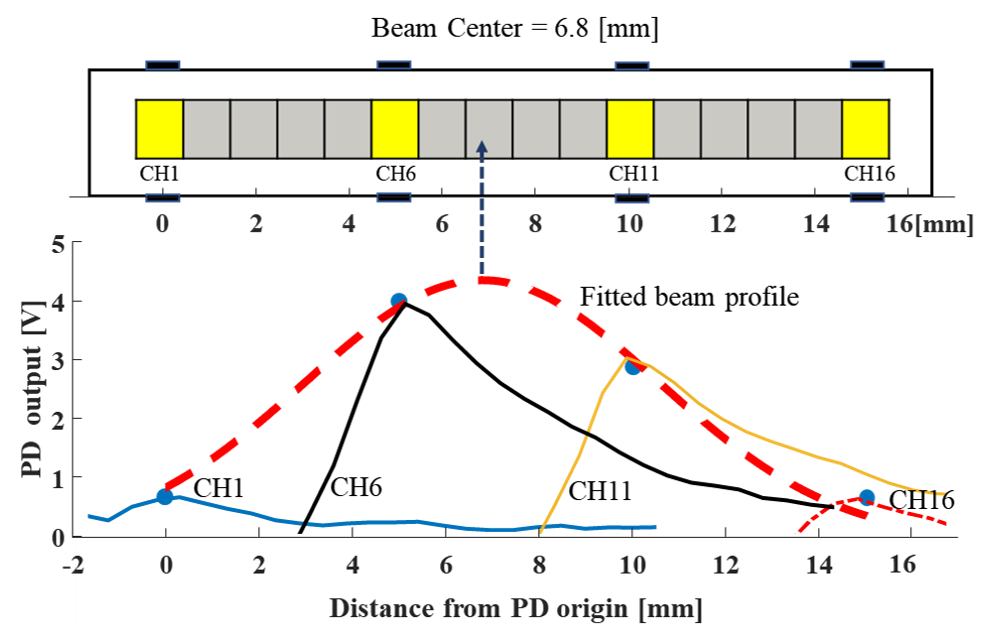}
    \caption{ Position of the beam center is estimated by applying a Gaussian fit on the signals of each diode channel. 
    }\label{Fig_PD_H_gauss}
    \end{figure}

From the three beams on the PD, the feature beam spot is selected as the closest one to the PD center. The center position for each beam can be estimated by analyzing the output signals in the spatial domain. 
Starting at the origin of the PD coordinate frame on $CH1$, the center and the end ($CH16$) of the PD would be 7.5 mm and 15 mm, respectively. On this PD coordinate frame, the output of each channel for a beam can be distributed as plotted in Fig. \ref{Fig_PD_H_gauss}.

The laser power spectrum was modeled as a Gaussian distribution\cite{laconte}, the beam center is estimated by applying a Gaussian fit on the signals of each PD channel. 
This paper used the iterative procedure of Gaussian fitting proposed in \cite{guo2011}, which is known to be a robust and fast method. 
If the natural log is applied on both sides of 1D Gaussian model, then a function of the second-order polynomial can be derived. 


  
    \begin{equation}
    \begin{array}{l}
        \ln (y) = \ln (A) - \frac{{{{(x - \mu )}^2}}}{{2{\sigma ^2}}}\\
        =  - \frac{{{x^2}}}{{2{\sigma ^2}}} + \frac{{2\mu x}}{{2{\sigma ^2}}} + \left( { - \frac{{{\mu ^2}}}{{2{\sigma ^2}}} + \ln (A)} \right) = {a_2}{x^2} + {a_1}x + {a_0}
    \end{array}
    \end{equation}
where x is the beam power input, y is the voltage output of the PD, $\mu$ is the average, or the beam center position, $\sigma$ is the standard deviation and A is a weight value. 


    
However, the actual output voltage signal, $y$, is corrupted with sensor noise.  To apply the iterative procedure of Gaussian fitting \cite{guo2011} to achieve more robustness to the noise, the equation is modified as

    \begin{equation}
    \begin{array}{l}
        \left[ {\begin{array}{*{20}{c}}
        {\sum\nolimits_{i = 1}^m {x_i^4y_{i,(k - 1)}^2} }&{\sum\nolimits_{i = 1}^m {x_i^3y_{i,(k - 1)}^2} }&{\sum\nolimits_{i = 1}^m {x_i^2y_{i,(k - 1)}^2} }\\
        {\sum\nolimits_{i = 1}^m {x_i^3y_{i,(k - 1)}^2} }&{\sum\nolimits_{i = 1}^m {x_i^2y_{i,(k - 1)}^2} }&{\sum\nolimits_{i = 1}^m {{x_i}y_{i,(k - 1)}^2} }\\
        {\sum\nolimits_{i = 1}^m {x_i^2y_{i,(k - 1)}^2} }&{\sum\nolimits_{i = 1}^m {{x_i}y_{i,(k - 1)}^2} }&{\sum\nolimits_{i = 1}^m {y_{i,(k - 1)}^2} }
        \end{array}} \right]*\\
        \left\{ {\begin{array}{*{20}{c}}
        {{a_{2,(k)}}}\\
        {{a_{1,(k)}}}\\
        {{a_{0,(k)}}}
        \end{array}} \right\} = \left\{ {\begin{array}{*{20}{c}}
        {\sum\nolimits_{i = 1}^m {x_i^2y_{i,(k - 1)}^2\ln ({y_i})} }\\
        {\sum\nolimits_{i = 1}^m {{x_i}y_{i,(k - 1)}^2\ln ({y_i})} }\\
        {\sum\nolimits_{i = 1}^m {y_{i,(k - 1)}^2\ln ({y_i})} }
        \end{array}} \right\}
    \end{array}
    \end{equation}
    
where k is the number of iterations, with a value of k=10 used in this study, and $y_{i(k)}$ is defined as
    
    \begin{equation}
        {y_{i,(k)}} = \left\{ {\begin{array}{*{20}{c}}
        {\,\,\,\,\,\,\,\,\,\,\,\,\,\,\,\,\,\,\,\,\,\,\,\,\,\,{y_i}\,\,\,\,\,\,\,\,\,\,\,\,\,\,\,\,\,\,\,\,\,\,\,\,\,\,\,\,\,\,\,\,\,\,\,\,\,\,\,\,\,\,\,\,\,for\,\,k = 0}\\
        {\exp \left( {{a_{2,(k)}}x_i^2 + {a_{1,(k)}}{x_i} + {a_{0,(k)}}} \right)for\,\,k > 0}
        \end{array}} \right.
    \end{equation}

Finally, the model parameters$\mu$ and $\sigma$ are then solved as 
    \begin{equation}
    {\sigma ^2} =  - \frac{1}{{2{a_{2,(k)}}}},\,\,\,\,\,\,\,\,\mu  = {a_{1,(k)}}{\sigma ^2} =  - \frac{{{a_{1,(k)}}}}{{2{a_{2,(k)}}}}
    \end{equation}

Since there were only four dataset measurements from the four laser scan channels, two more sets at -5 mm and 20 mm were augmented with the value of 0.1 V, which is approximately the noise level.

After estimating the center position for all the beams on the PD, the feature beam point is then selected that has the closest value to the center or 7.5 mm.

The next process is to match the corresponding beam point (${}^LP_i$) from the LiDAR raw data that contains the 3D position of each beam in polar or Cartesian coordinates with respect to the sensor body frame.
To find the correspondence beam, the photodetector needs to be detected and segmented from the target surface by LiDAR measurement. However, using only the 3D point data, the PD is not distinguishable from the target surface because the PD is too flat to show any clear depth differences.

Instead of using the LiDAR 3D data, this paper used the difference in the reflectivity values of the beam on the PD from their neighbor beams. 
The surroundings of the PD are covered with a black colored background to force a lower reflectance intensity than that on the PD sensor surface.  

\begin{figure}[!t]
    \centering
    \includegraphics[width=1\columnwidth]{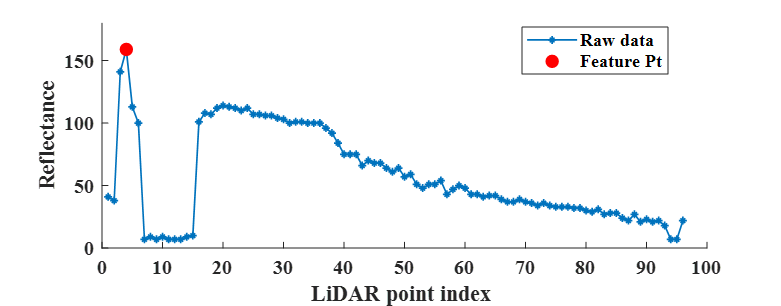}
    \caption{ Reflectivity values of the beams in the same laser channel that scan over the photodetector. The beam point with the local maxima intensity near the edges is the feature point on the photodetector.
    }\label{Fig_PD_H_reflectance}
    \end{figure}

The test results showed that the reflectivity of the beams that are projecting on the PD area have relatively higher values than those of the other nearby beams, as shown in Fig. \ref{Fig_PD_H_reflectance}.

Now, we have the matching pair of the position data of each key beam point, ${}^LP_i$($\omega$,$\alpha$,$r$) and ${}^OP_i($x,y,z$)$, which have been obtained from the PD signals and the LiDAR data, respectively.

\begin{figure}[!t]
    \centering
    \includegraphics[width=1\columnwidth]{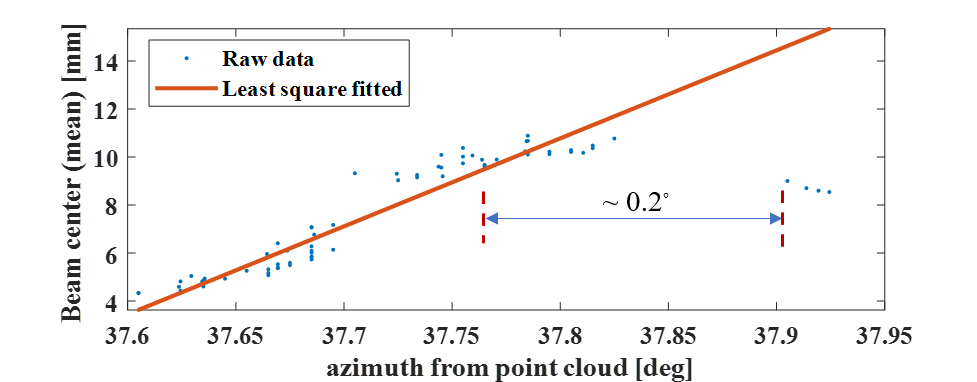}
    \caption{The azimuth ($\alpha$) from the point cloud data and the position of the beam center from the photodetector are plotted for 50 scans. The outliers are removed to obtain a linear relationship for correct correspondence matching.
    }\label{Fig_PD_H_postprocessing}
    \end{figure}

The azimuth ($\alpha$) from the point cloud and the matching beam center from the Gaussian fit is plotted for 50 LiDAR scans to validate the correspondence matching. Given that the azimuth of the key beam point varies over the scans due to the LiDAR rotation fluctuations, the beam center should be changed proportionally. However, the results showed some outliers deviated from the expected linear relationship, as plotted in Fig. \ref{Fig_PD_H_postprocessing}. 
All outliers have the offset error of 0.2 degrees, which is the azimuth resolution of one beam index in the same vertical channel. The offset by one index could have been caused due to several factors, such as the potential for data loss of the first or the last beam of the target surface by the sharp edge of the target, which occasionally occurs.

At the experimental configuration of the target position, such an index offset can cause as high as an 9.7 mm center position error. 

The actual relationship between the azimuth and the center position should be a function of the tangent, but we approximated it as a linear curve within a small range of 1 azimuth degree. The outliers removed using RANSAC are shown as the solid red curve in Fig. \ref{Fig_PD_H_postprocessing}. 
Thus, the pose estimation applied the new ${}^OP_x$ from this post-processing with all LiDAR scan data as

    \begin{equation}
        {}^D{{\bf{p}}_{i,k}} = \left[ {\begin{array}{*{20}{c}}
        {{{\widehat \mu }_{i,k}} = {\nu _i} + {\tau _i}{\alpha _{i,k}}}\\
        0\\
        {{}^D{z_{i,k}}}
        \end{array}} \right]
    \end{equation}

\subsection{Optimizing pose estimation}
Now, the corresponding feature points have been obtained from the target board and the LiDAR, and the relative pose of the LiDAR with respect to the target body frame can be estimated. 
This paper applied the Levenberg Marquardt method for the iterative optimization to minimize the cost function of the pose error to obtain the best fit values for the pose variables $\beta ^*$.

    \begin{equation}
        \begin{array}{l}
        {\beta ^*} = \arg {\min _\beta }\sum {\left\| {{}^{\bf{O}}{\bf{P}} - [{}_{\bf{L}}^{\bf{O}}{\bf{R}}|{}_{\bf{L}}^{\bf{O}}{\bf{T}}]{}^{\bf{L}}{\bf{P}}} \right\|} \\
        = \beta  - \eta {({{\bf{J}}^{\bf{T}}}{\bf{J}} + \lambda {\rm{diag}}({\bf{J}}))^{ - 1}}{{\bf{J}}^{\bf{T}}}{\bf{F}}(\beta ))\\
        \end{array}
    \end{equation}
where    
    \begin{equation*}
       \beta  = (\phi ,\theta ,\psi ,\Delta x,\Delta y,\Delta z)
    \end{equation*}

Here, $\bf{J}$ is the Jacobian matrix for the cost function $\bf{F}$ in Eq.\ref{eq_Fi}, a constant $\eta$=0.02, and the varying tuning rate $\lambda$ is initially set at 0.3.

    \begin{figure*}[!t]
    \normalsize
    \begin{equation}
    F{(\beta )_i} = \left[ {\begin{array}{*{20}{c}}
    {{}^O{P_{1,i}} + {r_i}(c{\alpha _i}c{\omega _i})(c\psi s\phi  - c\phi s\psi s\theta ) - {r_i}(s{\alpha _i})(s\psi s\phi  + c\psi c\phi s\theta ) - {r_i}(c{\alpha _i}s{\omega _i})(c\theta c\phi ) - \Delta x}\\
    {{r_i}(s{\alpha _i})(c\phi s\psi  - c\psi s\theta s\phi ) - {r_i}(c{\alpha _i}c{\omega _i})(c\psi c\phi  + s\psi s\theta s\phi ) - {r_i}(c{\alpha _i}s{\omega _i})c\theta s\phi  - \Delta y}\\
    {{}^O{P_{3,i}} + {r_i}(c{\alpha _i}s{\omega _i})s\theta  - {r_i}(s{\alpha _i})c\psi c\theta  - {r_i}(c{\alpha _i}c{\omega _i})c\theta s\psi  - \Delta z}
    \end{array}} \right]
    \label{eq_Fi}
    \end{equation}
    \hrulefill
    \end{figure*}

\section{Experiment}
\begin{figure}[!t]
    \centering
    \includegraphics[width=1\columnwidth]{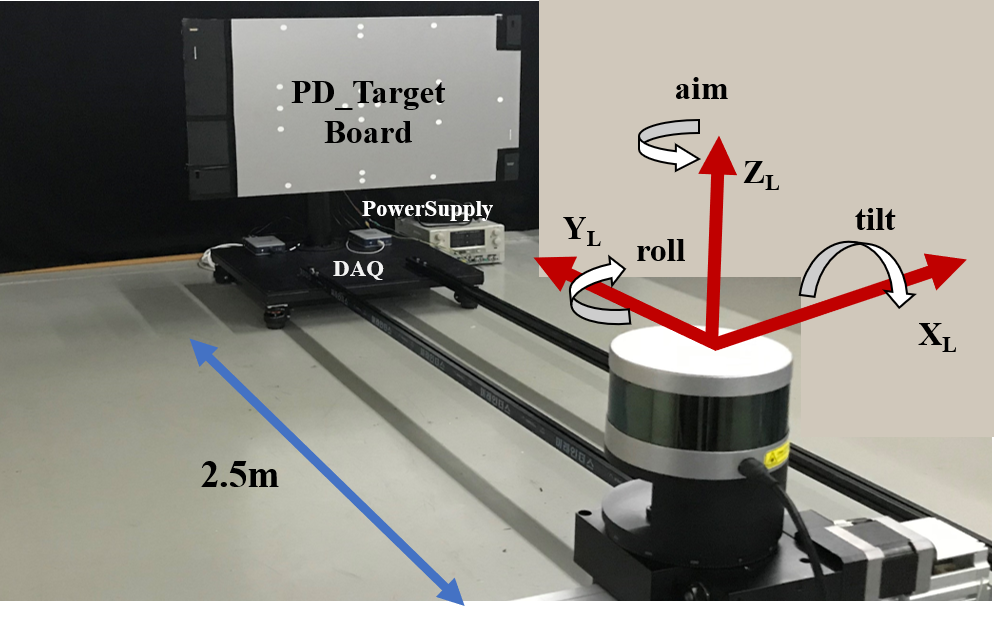}
    \caption{Experimental setup consisting of the PD-target board and the 2-DOF motion stage to control the LiDAR reference pose within a range of 3 degrees and 30 mm.
    }\label{Fig_ExpSetup}
    \end{figure}

Experiments were conducted to evaluate the accuracy and precision of the proposed system with the designed test bench that consisted of the PD-target board and the LiDAR pose controller module, as shown in Fig. \ref{Fig_ExpSetup}.

\subsection{Test-bench Setup}
The target is a flat aluminum board with dimensions of 1 m in width and 0.54 m in height and has 1D PD arrays attached near each board corner. The board surface surrounding the photodiodes was covered with black papers to reduce the surface reflectance such that it was lower than the photodiode surface. The photodiodes were aligned vertically on the top-left and bottom-right corners and aligned horizontally on the other remaining corners.
Two external DAQs of 4 ADC ports at a maximum of 4 MHz were used to capture the photodiode output voltage.
Note that only two photodetectors were used at a time due to limited availability of the high-performance ADCs. 

The LiDAR (VLP-16) is placed on the 2-DOF pose controller module positioned at 2.5 meters from the PD-target board. 
The pose controller module consisting of a rotation motor stage and a linear motor stage can control the precise yaw and x-position of the LiDAR. The yaw angle is controlled within 3 degrees of the range with a precision of 0.01 degree, and the horizontal position is controlled within 30 mm of the range with a precision of 0.01 mm.

\subsection{Experimental Results}
The initial pose of the LiDAR with respect to the target body frame is set as $X_O$=-0.7 m, $Y_O=-2.5$ m.
The relative pose of the LiDAR from the target board was estimated by the proposed algorithm at various reference poses of the yaw angle and x-position, 

\subsubsection{Yaw estimation test}
In the first test, the yaw rotation of the LiDAR was controlled from -3 to 3 degrees with a step of 0.5 degrees by the motor stage. For each reference yaw angle, 50 LiDAR scans were repeated to evaluate the accuracy and precision of the relative pose estimation.

\begin{figure}[!t]
    \centering
    \includegraphics[width=1\columnwidth]{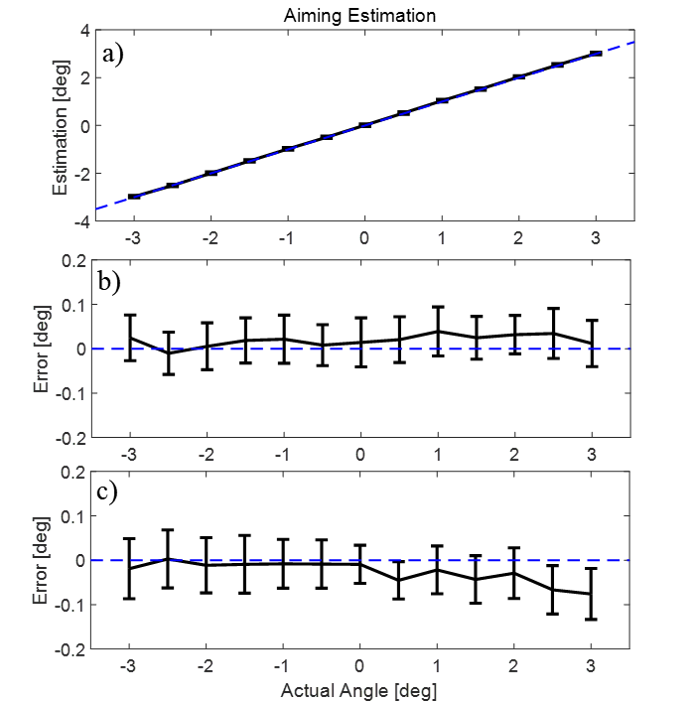}
    \caption{The yaw estimation a) tracking and error from -3 to 3 degrees b) for horizontal aligned and c) for vertical aligned PD
    }\label{Fig_PD_H_case12}
    \end{figure}

Fig. \ref{Fig_PD_H_case12}a) shows the horizontally aligned PD-target system is tracking the input yaw angle with high accuracy. 
The yaw estimation error plotted in Fig. \ref{Fig_PD_H_case12} b) indicates that the highest offset error is approximately 0.2 degrees at the reference angle of -3 degrees. The overall accuracy and precision of the yaw estimation are calculated to be 0.05 degrees. 

The same test was repeated with the vertically aligned photodetector array and plotted in Fig. \ref{Fig_PD_H_case12}c). The overall accuracy is under 0.03 degrees and the precision is under 0.06 degrees. The highest offset error is shown at the angle of 3 degrees, with the error value of approximately 0.1 degree. 




\subsubsection{Displacement estimation test}
\begin{figure}[!t]
    \centering
    \includegraphics[width=1\columnwidth]{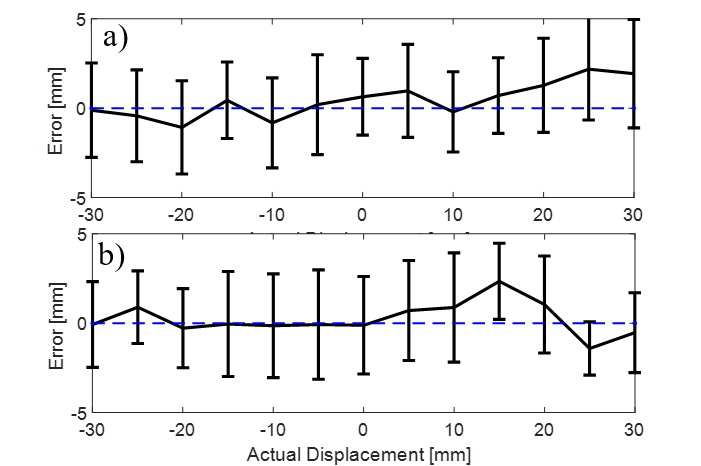}
    \caption{ The displacement estimation error statistics from 50 scans by a) Horizontal aligned b) vertically aligned photodetector. 
    }\label{Fig_PD_HV_case22}  
    \end{figure}

A similar test was conducted by varying the X-position of the LiDAR from -30 mm to 30 mm with a step of 5 mm. Through the test, all orientation angles were fixed at the initial values.
The estimation errors of the sensor position for both the horizontally and vertically aligned photodetectors are plotted in  Fig. \ref{Fig_PD_HV_case22}. 
As shown in the results, both alignment types can detect the LiDAR offset displacement with an error of less than 3 mm. 

\begin{figure}[!t]
    \centering
    \includegraphics[width=1\columnwidth]{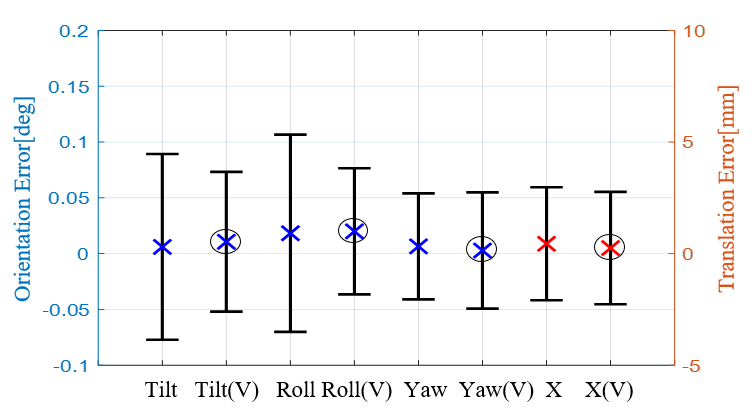}
    \caption{ Statistics of orientation angles and X-position errors for translation experiment. The circle mark indicates the results of the vertically aligned photodetectors.
    }\label{Fig_PD_comparison_case22}
    \end{figure}

Fig. \ref{Fig_PD_comparison_case22} shows the statistics of the estimation error for all orientations and the displacement for this test. 
The horizontally and vertically aligned photodetectors have  similar accuracy and precision, except the precision of the roll and tilt for the horizontal alignment seems slightly higher by  0.02 degrees. 

The overall evaluation of the proposed system for horizontal and vertical aligned photodetector from averaging all test results as presented in Table \ref{tab:exp}. Considering the possibility of experimental error, both photodetector alignments are observed to yield a similar level of accuracy and precision for all orientations and displacements, which are within 0.1 degree and 3 mm. 
Compared to the high depth measurement noise and low resolution of a mobile LiDAR, the 0.1 degree and 3 mm calibration errors are quite an improvement. If more photodetectors could be used, then the performance of the pose estimation would be improved more. 

    \begin{table}[t]
    \caption{ Overall Experimental Result of Alignment Estimation}
    \label{tab:exp}
    \begin{tabular}{cc|cccc}
    \hline
    \multicolumn{2}{c|}{Estimation} & Tilt $(^{\circ})$&	Roll $(^{\circ})$&Yaw $(^{\circ})$ &  $\Delta X (mm)$\\
    \hline
    \multirow{2}{*}{Horizontal PD} & Accuracy             & 0.03 & 0.04 & 0.01 & 0.6 \\
                                   & Precision            & 0.11 & 0.10 & 0.05 & 2.3 \\ \hline
    \multirow{2}{*}{Vertical PD}   & Accuracy             & 0.01 & 0.01 & 0.02 & 0.3 \\
                                   & Precision            & 0.08 & 0.08 & 0.05 & 2.6 \\
    \hline
    \end{tabular}
    \end{table}

\section{Conclusion}
This paper proposed a novel PD-target calibration system for the extrinsic parameter calibration of LiDAR mounted on a mobile platform for automatic sensor misalignment inspection.
The PD-target calibration system is composed of a planar target board with modules of NIR photodetector arrays attached near the corners of the board. With the photodetector arrays, the precise position of the laser beams on the board surface can be measured and used as the correspondence feature points for estimating the relative pose between the LiDAR and the target frame body. 
The proposed system was evaluated in terms of the accuracy and precision of the pose estimation with the designed test bench that can control the reference yaw and horizontal position of the LiDAR. 
Experimental results tested on VLP-16 LiDAR showed that the system estimated the offset pose within 0.15 degrees and 3 mm of accuracy and precision combined. 

There exist some technical issues that need to be overcome.
The projection of laser beams could depart from the photodiode  due to the low vertical resolution of LiDAR. 
Also, the cost of high speed ADC DAQ boards with many channels is another limtation and this study only could test two photodetectors at a time. More implementations of photodetectors would improve the overall performance. 

\bibliographystyle{IEEEtranTIE}
\bibliography{arxiv_PD_Target_shortend}\ 

\begin{thebibliography}{10}
\providecommand{\url}[1]{#1}
\csname url@samestyle\endcsname
\providecommand{\newblock}{\relax}
\providecommand{\bibinfo}[2]{#2}
\providecommand{\BIBentrySTDinterwordspacing}{\spaceskip=0pt\relax}
\providecommand{\BIBentryALTinterwordstretchfactor}{4}
\providecommand{\BIBentryALTinterwordspacing}{\spaceskip=\fontdimen2\font plus
\BIBentryALTinterwordstretchfactor\fontdimen3\font minus
  \fontdimen4\font\relax}
\providecommand{\BIBforeignlanguage}[2]{{%
\expandafter\ifx\csname l@#1\endcsname\relax
\typeout{** WARNING: IEEEtran.bst: No hyphenation pattern has been}%
\typeout{** loaded for the language `#1'. Using the pattern for}%
\typeout{** the default language instead.}%
\else
\language=\csname l@#1\endcsname
\fi
#2}}
\providecommand{\BIBdecl}{\relax}
\BIBdecl

\bibitem{Song2016}
H.~Song, W.~Choi, and H.~Kim, ``{Robust Vision-Based Relative-Localization
  Approach Using an RGB-Depth Camera and LiDAR Sensor Fusion},'' \emph{IEEE
  Transactions on Industrial Electronics}, vol.~63,
  \href{http://dx.doi.org/10.1109/TIE.2016.2521346}{DOI
  10.1109/TIE.2016.2521346}, no.~6, pp. 3725--3736, 2016.

\bibitem{Kang2012}
Y.~Kang, C.~Roh, S.~B. Suh, and B.~Song, ``{A lidar-based decision-making
  method for road boundary detection using multiple Kalman filters},''
  \emph{IEEE Transactions on Industrial Electronics}, vol.~59,
  \href{http://dx.doi.org/10.1109/TIE.2012.2185013}{DOI
  10.1109/TIE.2012.2185013}, no.~11, pp. 4360--4368, 2012.

\bibitem{Ramasamy2016}
\BIBentryALTinterwordspacing
S.~Ramasamy, R.~Sabatini, A.~Gardi, and J.~Liu, ``{LIDAR obstacle warning and
  avoidance system for unmanned aerial vehicle sense-and-avoid},''
  \emph{Aerospace Science and Technology}, vol.~55,
  \href{http://dx.doi.org/10.1016/j.ast.2016.05.020}{DOI
  10.1016/j.ast.2016.05.020}, pp. 344--358, 2016. [Online]. Available:
  \url{http://dx.doi.org/10.1016/j.ast.2016.05.020}
\BIBentrySTDinterwordspacing

\bibitem{Zeng2018}
Y.~Zeng, Y.~Hu, S.~Liu, J.~Ye, Y.~Han, X.~Li, and N.~Sun, ``{RT3D: Real-time
  3-D vehicle detection in LiDAR point cloud for autonomous driving},''
  \emph{IEEE Robotics and Automation Letters}, vol.~3,
  \href{http://dx.doi.org/10.1109/LRA.2018.2852843}{DOI
  10.1109/LRA.2018.2852843}, no.~4, pp. 3434--3440, 2018.

\bibitem{Zhang2020}
Y.~Zhang, L.~Chen, Z.~XuanYuan, and W.~Tian, ``{Three-Dimensional Cooperative
  Mapping for Connected and Automated Vehicles},'' \emph{IEEE Transactions on
  Industrial Electronics}, vol.~67,
  \href{http://dx.doi.org/10.1109/TIE.2019.2931521}{DOI
  10.1109/TIE.2019.2931521}, no.~8, pp. 6649--6658, 2020.

\bibitem{Wang2017}
\BIBentryALTinterwordspacing
L.~Wang, Y.~Zhang, and J.~Wang, ``{Map-Based Localization Method for Autonomous
  Vehicles Using 3D-LIDAR},'' \emph{IFAC-PapersOnLine}, vol.~50,
  \href{http://dx.doi.org/10.1016/j.ifacol.2017.08.046}{DOI
  10.1016/j.ifacol.2017.08.046}, no.~1, pp. 276--281, 2017. [Online].
  Available: \url{https://doi.org/10.1016/j.ifacol.2017.08.046}
\BIBentrySTDinterwordspacing

\bibitem{Zhu2013}
Z.~Zhu and J.~Liu, ``{Unsupervised Extrinsic Parameters Calibration for
  Multi-beam LIDARs},'' \href{http://dx.doi.org/10.2991/iccsee.2013.278}{DOI
  10.2991/iccsee.2013.278}, no. Iccsee, pp. 1110--1113, 2013.

\bibitem{Zaiter2019}
M.~A. Zaiter, R.~Lherbier, G.~Faour, O.~Bazzi, and J.~C. Noyer, ``{3D LiDAR
  extrinsic calibration method using ground plane model estimation},''
  \emph{2019 8th IEEE International Conference on Connected Vehicles and Expo,
  ICCVE 2019 - Proceedings},
  \href{http://dx.doi.org/10.1109/ICCVE45908.2019.8964949}{DOI
  10.1109/ICCVE45908.2019.8964949}, pp. 1--6, 2019.

\bibitem{Atanacio-Jimenez2011}
G.~Atanacio-Jim{\'{e}}nez, J.~J. Gonz{\'{a}}lez-Barbosa, J.~B. Hurtado-Ramos,
  F.~J. Ornelas-Rodr{\'{i}}guez, H.~Jim{\'{e}}nez-Hern{\'{a}}ndez,
  T.~Garc{\'{i}}a-Ramirez, and R.~Gonz{\'{a}}lez-Barbosa, ``{LIDAR velodyne
  HDL-64E calibration using pattern planes},'' \emph{International Journal of
  Advanced Robotic Systems}, vol.~8, \href{http://dx.doi.org/10.5772/50900}{DOI
  10.5772/50900}, no.~5, pp. 70--82, 2011.

\bibitem{Muhammad2010}
N.~Muhammad and S.~Lacroix, ``{Calibration of a rotating multi-beam Lidar},''
  \emph{IEEE/RSJ 2010 International Conference on Intelligent Robots and
  Systems, IROS 2010 - Conference Proceedings},
  \href{http://dx.doi.org/10.1109/IROS.2010.5651382}{DOI
  10.1109/IROS.2010.5651382}, pp. 5648--5653, 2010.

\bibitem{Glennie2010}
C.~Glennie and D.~D. Lichti, ``{Static calibration and analysis of the velodyne
  HDL-64E S2 for high accuracy mobile scanning},'' \emph{Remote Sensing},
  vol.~2, \href{http://dx.doi.org/10.3390/rs2061610}{DOI 10.3390/rs2061610},
  no.~6, pp. 1610--1624, 2010.

\bibitem{geiger2012automatic}
A.~Geiger, F.~Moosmann, {\"O}.~Car, and B.~Schuster, ``Automatic camera and
  range sensor calibration using a single shot,'' in \emph{2012 IEEE
  International Conference on Robotics and Automation}, pp. 3936--3943.\hskip
  1em plus 0.5em minus 0.4em\relax IEEE, 2012.

\bibitem{kim2019extrinsic}
E.-S. Kim and S.-Y. Park, ``Extrinsic calibration of a camera-lidar multi
  sensor system using a planar chessboard,'' in \emph{2019 Eleventh
  International Conference on Ubiquitous and Future Networks (ICUFN)}, pp.
  89--91.\hskip 1em plus 0.5em minus 0.4em\relax IEEE, 2019.

\bibitem{fremont2008extrinsic}
V.~Fremont, P.~Bonnifait \emph{et~al.}, ``Extrinsic calibration between a
  multi-layer lidar and a camera,'' in \emph{2008 IEEE International Conference
  on Multisensor Fusion and Integration for Intelligent Systems}, pp.
  214--219.\hskip 1em plus 0.5em minus 0.4em\relax IEEE, 2008.

\bibitem{alismail2012automatic}
H.~Alismail, L.~D. Baker, and B.~Browning, ``Automatic calibration of a range
  sensor and camera system,'' in \emph{2012 Second International Conference on
  3D Imaging, Modeling, Processing, Visualization \& Transmission}, pp.
  286--292.\hskip 1em plus 0.5em minus 0.4em\relax IEEE, 2012.

\bibitem{gong20133d}
X.~Gong, Y.~Lin, and J.~Liu, ``3d lidar-camera extrinsic calibration using an
  arbitrary trihedron,'' \emph{Sensors}, vol.~13, no.~2, pp. 1902--1918, 2013.

\bibitem{pusztai2017accurate}
Z.~Pusztai and L.~Hajder, ``Accurate calibration of lidar-camera systems using
  ordinary boxes,'' in \emph{Proceedings of the IEEE International Conference
  on Computer Vision Workshops}, pp. 394--402, 2017.

\bibitem{zhang2004extrinsic}
Q.~Zhang and R.~Pless, ``Extrinsic calibration of a camera and laser range
  finder (improves camera calibration),'' in \emph{2004 IEEE/RSJ International
  Conference on Intelligent Robots and Systems (IROS)(IEEE Cat. No.
  04CH37566)}, vol.~3, pp. 2301--2306.\hskip 1em plus 0.5em minus 0.4em\relax
  IEEE, 2004.

\bibitem{verma2019automatic}
S.~Verma, J.~S. Berrio, S.~Worrall, and E.~Nebot, ``Automatic extrinsic
  calibration between a camera and a 3d lidar using 3d point and plane
  correspondences,'' in \emph{2019 IEEE Intelligent Transportation Systems
  Conference (ITSC)}, pp. 3906--3912.\hskip 1em plus 0.5em minus 0.4em\relax
  IEEE, 2019.

\bibitem{Mirzaei2012}
F.~M. Mirzaei, D.~G. Kottas, and S.~I. Roumeliotis, ``{3D LIDAR-camera
  intrinsic and extrinsic calibration: Identifiability and analytical
  least-squares-based initialization},'' \emph{International Journal of
  Robotics Research}, vol.~31,
  \href{http://dx.doi.org/10.1177/0278364911435689}{DOI
  10.1177/0278364911435689}, no.~4, pp. 452--467, 2012.

\bibitem{kim2012developing}
Y.-K. Kim, Y.~Kim, Y.~S. Jung, I.~G. Jang, K.-S. Kim, S.~Kim, and B.~M. Kwak,
  ``Developing accurate long-distance 6-dof motion detection with
  one-dimensional laser sensors: Three-beam detection system,'' \emph{IEEE
  Transactions on Industrial Electronics}, vol.~60, no.~8, pp. 3386--3395,
  2012.

\bibitem{kim2014design}
Y.-K. Kim, I.~G. Jang, K.-S. Kim, and S.~Kim, ``Design improvement of the
  three-beam detector towards a precise long-range 6-degree of freedom motion
  sensor system,'' \emph{Review of Scientific Instruments}, vol.~85, no.~1, p.
  015004, 2014.

\bibitem{kim2015portable}
Y.-K. Kim, K.-S. Kim, and S.~Kim, ``A portable and remote 6-dof pose sensor
  system with a long measurement range based on 1-d laser sensors,'' \emph{IEEE
  Transactions on Industrial Electronics}, vol.~62, no.~9, pp. 5722--5729,
  2015.

\bibitem{kim2014structural}
Y.-K. Kim, I.~G. Jang, Y.~Kim, K.-S. Kim, and S.~Kim, ``Structural optimization
  of a novel 6-dof pose sensor system for enhancing noise robustness at a long
  distance,'' \emph{IEEE Transactions on Industrial Electronics}, vol.~61,
  no.~10, pp. 5622--5631, 2014.

\bibitem{kim2014note}
Y.-K. Kim and K.-S. Kim, ``Note: Reliable and non-contact 6d motion tracking
  system based on 2d laser scanners for cargo transportation,'' \emph{Review of
  Scientific Instruments}, vol.~85, no.~10, p. 106102, 2014.

\bibitem{Underwood2010}
\BIBentryALTinterwordspacing
J.~P. Underwood, A.~Hill, T.~Peynot, and S.~J. Scheding, ``{Error modeling and
  calibration of exteroceptive sensors for accurate mapping applications},''
  \emph{Journal of Field Robotics}, vol.~27,
  \href{http://dx.doi.org/10.1002/rob.20315}{DOI 10.1002/rob.20315}, no.~1, pp.
  2--20, 2010. [Online]. Available: \url{https://doi.org/10.1002/rob.20315}
\BIBentrySTDinterwordspacing

\bibitem{Fuada2017}
S.~Fuada, A.~P. Putra, Y.~Aska, and T.~Adiono, ``{Trans-impedance amplifier
  (HA) design for Visible Light Communication (VLC) using commercially
  available OP-AMP},'' \emph{Proceedings - 2016 3rd International Conference on
  Information Technology, Computer, and Electrical Engineering, ICITACEE 2016},
  \href{http://dx.doi.org/10.1109/ICITACEE.2016.7892405}{DOI
  10.1109/ICITACEE.2016.7892405}, no. Vlc, pp. 31--36, 2017.

\bibitem{laconte}
J.~{Laconte}, S.~{Deschênes}, M.~{Labussière}, and F.~{Pomerleau}, ``Lidar
  measurement bias estimation via return waveform modelling in a context of 3d
  mapping,'' in \emph{2019 International Conference on Robotics and Automation
  (ICRA)}, pp. 8100--8106, 2019.

\bibitem{guo2011}
H.~{Guo}, ``A simple algorithm for fitting a gaussian function [dsp tips and
  tricks],'' \emph{IEEE Signal Processing Magazine}, vol.~28, no.~5, pp.
  134--137, 2011.

\end{thebibliography}

\end{document}